\documentclass[english,12pt,aps,prd,a4paper,preprintnumbers,floatfix,nofootinbib,superscriptaddress, notitlepage]{revtex4-1} 
\usepackage[usenames,dvipsnames]{color}  %%%%%%% usenames,dvipsnames required to get BrickRed
\usepackage{graphicx}
\usepackage{setspace}
\usepackage{caption}
\usepackage{subcaption}
\usepackage{amsmath}
\usepackage{amssymb}
\usepackage[colorlinks=true,citecolor=darkred,urlcolor=darkred, pdfborder={0 0 0}]{hyperref}
\usepackage[normalem]{ulem}
\usepackage{xcolor}
\usepackage{array}
\usepackage{bm}

\makeatletter
\def\p@subsection{}
\makeatother

\definecolor{darkred}{rgb}{0.6,0,0}

\definecolor{linkcolor}{rgb}{0,0,0.5}

\def\gsim{\raise0.3ex\hbox{$\;>$\kern-0.75em\raise-1.1ex\hbox{$\sim\;$}}}
\def\lsim{\raise0.3ex\hbox{$\;<$\kern-0.75em\raise-1.1ex\hbox{$\sim\;$}}}

\def\beqn#1{\begin{equation}\label{#1}}
\def\eeqn{\end{equation}}

\def\beqa#1{\begin{eqnarray}\label{#1}}
\def\eeqa{\end{eqnarray}}

\def\Z2{$\mathcal{Z_2}$}

\def\hc{\mathrm{h.c.}}

\newcommand {\ignore}[1]{}

\def\SM{$\mathrm{SU(3)_c \otimes SU(2)_L \otimes U(1)_Y}$ }

\def\BSM{$\mathrm{SU(3)_c \otimes SU(2)_L \otimes U(1)_Y \otimes \mathbb{Z}_2}$ }
\def\321{$\mathrm{SU(3) \otimes SU(2) \otimes U(1)}$ }

\newcommand{\AddrIISERB}{Department of Physics, Indian Institute of Science Education and Research - Bhopal, \\ 
Bhopal Bypass Road, Bhauri, Bhopal 462066, India}

\begin{document}

\title{\color{BrickRed}W boson mass in Singlet-Triplet Scotogenic dark matter model}
\author{Aditya Batra}\email{adityab17@iiserb.ac.in}
\affiliation{\AddrIISERB}
\author{ShivaSankar K.A}\email{shivasankar17@iiserb.ac.in}
\affiliation{\AddrIISERB}
\author{Sanjoy Mandal}\email{smandal@kias.re.kr}
\affiliation{Korea Institute for Advanced Study, Seoul 02455, Korea}
\author{Rahul Srivastava}\email{rahul@iiserb.ac.in}
\affiliation{\AddrIISERB}

\begin{abstract}
\vspace{1cm} 
The recent high precision measurement of $W$ boson mass by CDF-II collaboration points to the contribution(s) of new physics beyond the Standard Model. One of the minimalistic ways to account for the anomalous  $W$ boson mass is by introducing a hyperchargeless real $SU(2)_L$ triplet scalar whose vacuum expectation value explicitly contributes to the $W$ boson mass at the tree level while the $Z$ boson mass remains the same. Such a triplet can be naturally embedded in a singlet-triplet scotogenic model for one loop neutrino mass generated by dark sector particles running in the loop. We discuss the detailed phenomenology  of the model, obtaining the parameter space consistent with the CDF-II $W$ boson mass measurements. The dark matter as well as the constraints comings from $S$, $T$, $U$ parameters are also analyzed.

\end{abstract}
%%%%%%%%%%%%%%%%%%%%%%%%%%%%
\maketitle

%%%%%%%%%%%%%%%%%%%%%%%%%%%%%%%%%%%%%%%%%%%%%%%%%%%%%%%%%%%%%%%%%%%%%%%%%%

\section{INTRODUCTION}\label{sec1-intro}
%%%%%%%%%%%%%%%%%%%%%%%%%%%%%%%%%%%%%%%%%%%%%%%%%%
Some of the main goals of Large Hadron Collider (LHC) have been to shed more light on the detailed mechanism of spontaneous symmetry breaking in the Standard Model (SM) and to look for possible new physics beyond Standard Model (BSM). The ATLAS~\cite{Aad:2012tfa} and CMS~\cite{Chatrchyan:2012xdj} experiments have partially fulfilled this aim with the discovery of a scalar particle of mass $125$ GeV with properties that are similar to those of the SM Higgs. The discovery of neutrino oscillations~\cite{Super-Kamiokande:1998kpq} provides us with another important milestone in particle physics.  At many scales, there has now been significant evidence for the presence of cosmic dark matter~\cite{Planck:2018vyg} whose deep understanding we still lack. Therefore, despite its outstanding achievements, it is now widely expected that the SM cannot be the final theory of nature.

Recently the CDF-II collaboration has reported their high precision measurement of $W$ boson mass $M_{W}^{\rm CDF-II} = 80.4335 \pm 0.0094 \; \text{GeV}$~\cite{CDF:2022hxs} which shows a 7-$\sigma$ deviation from the SM expectation~($M_W^{\rm SM} = 80.354 \pm 0.007 \; \text{GeV}$)~\cite{Zyla:2020zbs} . Such a strong deviation from the SM prediction opens up the possibilities of studying new physics contributions to $W$ boson mass~\cite{Fan:2022dck,Lu:2022bgw,Athron:2022qpo,Yuan:2022cpw,Strumia:2022qkt,Yang:2022gvz,deBlas:2022hdk,Zhu:2022tpr,Du:2022pbp,Tang:2022pxh,Cacciapaglia:2022xih,Blennow:2022yfm,Sakurai:2022hwh,Fan:2022yly,Liu:2022jdq,Lee:2022nqz,Cheng:2022jyi,Bagnaschi:2022whn,Paul:2022dds,Bahl:2022xzi,Asadi:2022xiy,DiLuzio:2022xns,Athron:2022isz,Gu:2022htv,Heckman:2022the,Babu:2022pdn,Zhu:2022scj,Balkin:2022glu,Biekotter:2022abc,Endo:2022kiw,Crivellin:2022fdf,Cheung:2022zsb,Du:2022brr,Heo:2022dey,Krasnikov:2022xsi,Ahn:2022xeq,Han:2022juu,Zheng:2022irz,Perez:2022uil,Ghoshal:2022vzo,Kawamura:2022uft,Nagao:2022oin,Kanemura:2022ahw,Mondal:2022xdy,Zhang:2022nnh,Borah:2022obi,Chowdhury:2022moc,Arcadi:2022dmt,Cirigliano:2022qdm,Bagnaschi:2022qhb,Popov:2022ldh,Bhaskar:2022vgk,Du:2022fqv,Ghorbani:2022vtv,Carpenter:2022oyg}.

Of particular interest are new physics models which can relate the CDF-II anomaly with the other shortcomings  of the SM such as the long standing open questions in neutrino and dark matter physics. In this article, we discuss such an extension which explains the CDF-II anomaly along with generating naturally small one loop masses for the neutrinos through a novel variant of the canonical scotogenic model~\cite{Ma:2006km}. The model contains a real $SU(2)_L$ triplet scalar whose vacuum expectation value (VEV) can modify the $W$ boson mass without changing $Z$ boson mass. We show that with only a triplet scalar one can explain the new CDF-II measurements at tree level itself. We further show that such a real triplet scalar is naturally embedded in a generalization of the scotogenic model initially proposed in ~\cite{Hirsch:2013ola,Avila:2019hhv}. This simple models contains just a few new fields beyond the SM particle content: one singlet fermion $N$ and one hyperchargeless $SU(2)_L$ triplet fermion $\Sigma$. These are odd under a $\mathbb{Z}_2$ dark symmetry. Moreover, we include two new scalars: one $\mathbb{Z}_2$ odd $SU(2)_L$ doublet $\eta$ and the $\mathbb{Z}_2$ even $SU(2)_L$ triplet scalar $\Omega$ carrying no hypercharge. The advantage with this model is that the lack of neutrino mass and of a viable WIMP dark matter candidate, have a common origin. In addition we show that along with the small induced scalar triplet VEV, one gets additional contribution to the $W$ boson mass through the loop quantum corrections quantified by the $S$, $T$ and $U$ parameters. 

The plan of the paper is as follows. 
In Section  \ref{sec2-Wmass} we present the modification of $W$ boson mass from a tree level interaction with a real scalar $\mathrm{SU(2)_L}$ triplet.
In Section~\ref{sec3-model} we summarize the main properties of the model and discuss the singlet-triplet scotogenic model for neutrino mass generation.  
In Section~\ref{sec4-STU} we discuss the $S$, $T$, $U$ parameter space for the model in consideration with CDF-II results and compare it with the experimental constraints.
In Section~\ref{sec5-const} we discuss the dark matter constraints for the case of the scalar dark matter. Finally, we summarize our results in Sec.~\ref{sec6-conclusions}.

%%%%%%%%%%%%%%%%%%%%%%%%%%%%%%%%%%%%%%%%%%%%%%%%%%%%%%%%%%%%%%%%%%%%%%%%%%%%%%%%
\section{Real triplet scalar contribution to the $W$ mass}
\label{sec2-Wmass}
%%%%%%%%%%%%%%%%%%%%%%%%%%%%%%%%%%%%%%%%%%%%%%%%%%%%%%%%%%%%%%%%%%%%%%%%%%%%%%%%%
Before discussing the singlet-triplet scotogenic model in details, in this section we briefly discuss the impact of adding a $SU(2)_L$ triplet scalar $\Omega$ with $U(1)_Y = 0$ to the particle content of the SM.   The new interaction term relevant for $W$ boson mass in the Lagrangian is given by
\begin{equation}\label{eq:lagO}
	\mathcal{L}_{\Omega} = \text{Tr} \left[ (D_{\mu} \Omega)^{\dagger} (D^{\mu} \Omega) \right], \,\,\,\,\text{with}\,\,\,\Omega = 
\begin{pmatrix}
	\frac{\Omega^0}{\sqrt{2}} & \Omega^+\\
		\Omega^{-}	&         -\frac{\Omega^0}{2} \\
\end{pmatrix}
\end{equation}
where the covariant derivative $D_{\mu}$ is defined as
\begin{equation}
	D_{\mu} \Omega= \partial_{\mu} \Omega + i g  \left[ W_{\mu},  \Omega\right]
\end{equation}
with $g$ being the weak coupling constant. After electroweak symmetry breaking both the $SU(2)_L$ doublet scalar $\Phi$ and triplet scalar $\Omega$ get VEVs
\begin{eqnarray}
\langle \phi^0 \rangle = \frac{v_\Phi}{\sqrt{2}} \, , \quad \langle \Omega^0 \rangle = v_\Omega
\end{eqnarray}

With $v_\Omega\neq 0$, we can calculate the contribution of the triplet scalar to masses of gauge bosons. The masses of $W$ and $Z$ boson are given by,
\begin{align}
M_{W} = \frac{g}{2} \sqrt{v_\Phi^2 + 4 v_{\Omega}^2}\,\,\text{and}\,\,
		M_{Z} = \frac{\sqrt{g^2 + g'^2}}{2} v_\Phi
		\label{eq:Wmass}
\end{align}
where $g'$ is the coupling constant associated with the $U(1)_Y$ gauge symmetry.  Clearly, since $\Omega$ has zero hypercharge, it only contributes to the $W$ boson mass. Accordingly, the $\rho$ parameter modifies to 
\begin{align}
	\rho  & = \frac{\sqrt{v_{\Phi}^2 + 4v^2_{\Omega}} }{v_{\Phi}} \approx 1 + 2 \frac{v_{\Omega}^2}{v_{\Phi}^2}
\end{align}

Since, now the mass of $W$ boson also depends on $v_{\Omega}$, the new CDF-II measurement can be used to put ``naive bounds'' on $v_{\Omega}$\footnote{This naive bound is just for illustrative purposes in order to discuss the idea in a simplified setup. Of course the CDF-II or for that matter any experimental mass measurement cannot be directly equated to tree level theoretical values. We do a more careful analysis taking into account the impact of loop corrections and the resulting oblique parameters in later sections.}. We can constrain $v_{\Omega}$ using the equation
\begin{equation}\label{eq:Wcontrib}
	M_{W,\text{CDF}}^2 - M_{W,\text{SM}}^2 = g^2 v_{\Omega}^2
\end{equation}
where $M_{W,\text{SM}}^2 = g^2 v_{\Phi}^2/4 $ is the tree level value of the $W$ boson mass in the SM.
%%%%%%%%%%%%%%%%%%
\begin{figure}[t]
\centering
\includegraphics[scale=1]{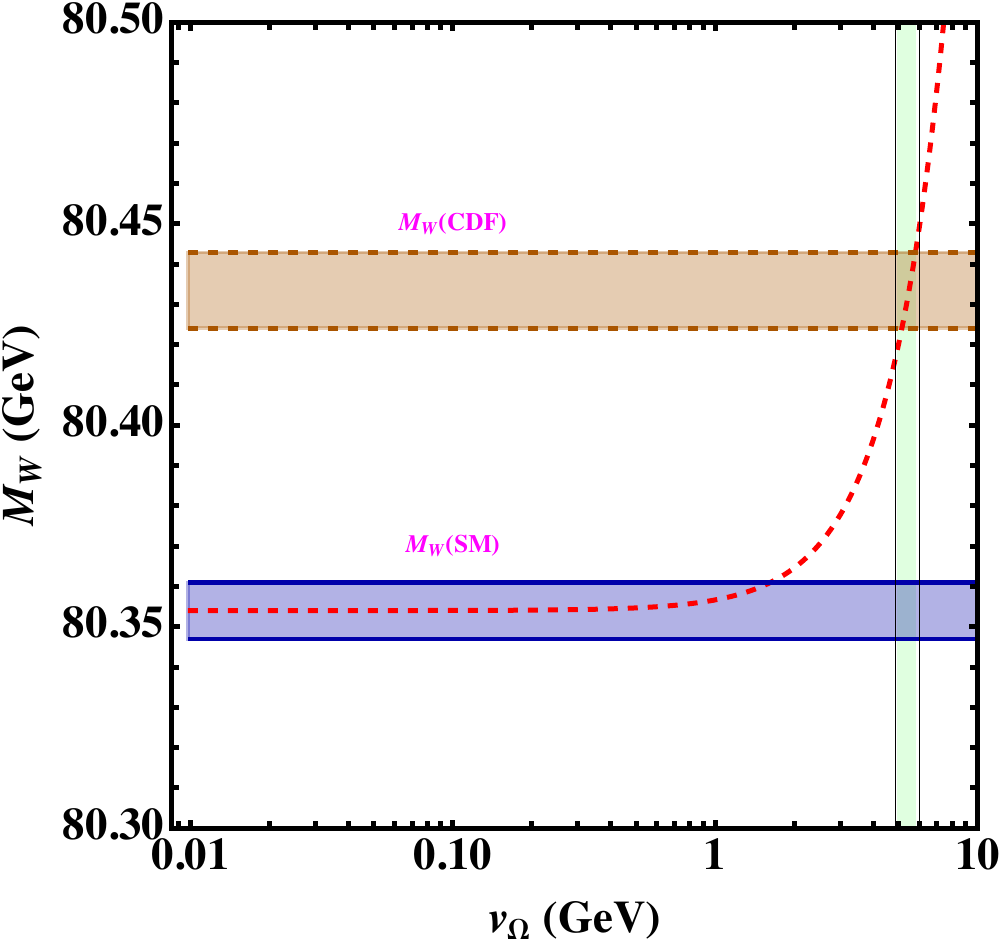}
\caption{The mass of $W$ boson vs VEV $v_{\Omega}$ of the triplet scalar. The horizontal purple band corresponds to the SM prediction~\cite{Zyla:2020zbs} while the dark orange band corresponds to the 1-$\sigma$ limit from the CDF-II measurement~\cite{CDF:2022hxs}. The vertical green band denotes the range of $v_\Omega$ which is compatible with the CDF-II result. }
\label{fig:Mvsvev}
\end{figure}
%%%%%%%%%%%%%%%%%%%%%%%%%

One can use \eqref{eq:Wcontrib} to plot the tree level change in $W$ boson mass with respect to the $v_{\Omega}$ as shown in Fig.~\ref{fig:Mvsvev}. From Fig.~\ref{fig:Mvsvev} we can see that the mass of the $W$ boson increases with the increase in the VEV  of the triplet scalar. The purple shaded region is the 1-$\sigma$ range of the $W$ boson mass given by the Particle Data Group~\cite{ParticleDataGroup:2020ssz}. The 1-$\sigma$ range of the new CDF-II measured is shown in dark orange color~\cite{CDF:2022hxs}. Considering the new CDF-II measurement and the SM prediction and requiring that the new $W$ mass contribution comes from the VEV of $\Omega^0$, the constraint in equation \eqref{eq:Wcontrib} translates to a bound on the VEV given by 
\begin{eqnarray}
 4.9 \;\text{GeV} \lessapprox v_{\Omega} \lessapprox 6.0\; \text{GeV}
\end{eqnarray}
 as indicated by the vertical green region in Fig.~\ref{fig:Mvsvev}. Thus, the addition of the hyperchargeless real triplet scalar can easily modify the mass of the $W$ boson consistent with CDF-II measured value without changing the $Z$ boson mass.
 However, it will be much more interesting to see if such a scalar can be naturally connected to other experimental and observational shortcomings of the SM in an intimate and self consistent way. In the rest of the paper we discuss such a scenario where the real triplet scalar is naturally embedded in a simple model for neutrino masses and dark matter.

%%%%%%%%%%%%%%%%%%%%%%%%%%%%%%%%%%%%%%%%%%%%%%%%%%%%%%%%%%%
\section{The Singlet Triplet Scotogenic Model}
\label{sec3-model}
%%%%%%%%%%%%%%%%%%%%%%%%%%%%%%%%%%%%%%%%%%%%%%%%%%%%%%%%%%%%%%%%%%
In this section we discuss the details of the singlet-triplet scotogenic model which generates small neutrino masses at one loop level with the dark matter running in the neutrino mass loop. The particle content of the model and their respective charges are given in Table \ref{tab1}.
\begin{table}[h]
\begin{center}
\begin{tabular}{| c| c || c | c |}
 \hline
\hspace{0.5cm}Fields \hspace{0.5cm}   & \hspace{0.5cm} \SM \hspace{0.5cm} & \hspace{0.5cm} $\mathbb{Z}_2$ \hspace{0.5cm} \\
\hline \hline
	$L$ \ \             & $(1, 2, -\frac{1}{2})$\ \      & $+1$ \\
	$ \ell$ \ \        & $(1, 1, -1)$\ \                    & $+1$ \\
	$\Phi$       \ \  & $(1, 2, \frac{1}{2})$\ \        & $+1$  \\
\hline \hline 
	$\Sigma$ \ \ 	  & $(1,3,0)$\ \						    &$-1$ \\
	$ N$ \ \           & $(1,1,0)$ \ \ 					    &$-1$ \\
\hline \hline 
	$\eta$			\ \ &$(1,2,\frac{1}{2})$ 				  \ \			&$-1$ \\
	$\Omega$ \ \  & $(1, 3, 0)$              \ \       &$-1$ \\
\hline 
  \end{tabular}
\end{center}
\caption{Particle content and  charge assignments for singlet-triplet scotogenic model. The flavour indices are suppressed for brevity.}
  \label{tab1}
\end{table} 
%%%%%%%%%%%%%%%%%%%%%%%%%%%%%%%%%%%%%%%%%%%%%%%%%%%%%%%%%%%%%%%%%%%%%
The extended particle content includes three generations of  singlet Majorana fermions $N$ and hyperchargeless triplet Majorana fermions $\Sigma$, both of which are charged odd under a $\mathbb{Z}_2$ dark symmetry. In the scalar sector, apart from the Higgs doublet $\Phi$, we have added an $SU(2)_L$ doublet scalar which is odd under $\mathbb{Z}_2$ and a real hyperchargeless $SU(2)_L$ triplet scalar $\Omega$ which is charged even under $\mathbb{Z}_2$.  Due to the $\mathbb{Z}_2$ symmetry conservation, the doublet scalar $\eta$ should not acquire any VEV, resulting in the neutrino masses arising from a loop diagram as shown in Fig.~\ref{fig:scotoloop}. As we have discussed, the VEV of the neutral component of the triplet scalar $\Omega$ contributes also to the $W$ boson mass. 

The \BSM invariant Yukawa Lagrangian for the particle content given in Table.~\ref{tab1} is
%%%%%%%%%%%%%%%%%%%%%%%%%%%%%%%%%%%%%%%%%%%%%%%%%%%%%%%%%%%%%%%%%%%%%%%%%
\begin{equation}
- \mathcal{L}_Y = Y_e^{\alpha \beta}\,\overline{L}_{\alpha} \, \Phi \, \ell_{\beta} + Y_{N}^\alpha \, \overline{L}_{\alpha} \, \tilde{\eta} \, N + Y_{\Sigma}^\alpha \, \overline{L}_{\alpha} \,C\Sigma^{\dagger}\, \tilde{\eta} \, + Y_{\Omega} \, \text{Tr}\big[\overline{\Sigma} \Omega\big] \, N + \frac{1}{2} \, M_\Sigma \, \text{Tr}\big[\overline{\Sigma}^{c} \Sigma\big]
+ \frac{1}{2} \, M_F \, \overline{N}^{c} N +  \hc 
\label{eq:yukawa}
\end{equation}
%%%%%%%%%%%%%%%%%%%%%%%%%%%%%%%%%%%%%%%%%%%%%%%%%%%%%%%%%%%
where $\alpha,\beta=1,2,3$ are the flavour indices and $\tilde{\eta}=i\sigma_2\eta^*$; $\sigma_2$ being the second Pauli matrix. For sake of brevity we have suppressed the flavour indices throughout this work unless explicitly stated.
The scalar sector potential invariant under the \BSM symmetry is given by
\begin{align}
\mathcal V &= -m_{\Phi}^2 \Phi^\dagger \Phi + m_{\eta}^2 \eta^\dagger \eta + \frac{\lambda_1}{2} \left( \Phi^\dagger \Phi \right)^2 + \frac{\lambda_2}{2} \left( \eta^\dagger \eta \right)^2 + \lambda_3 \left( \Phi^\dagger \Phi \right)\left( \eta^\dagger \eta \right) 
  + \lambda_4 \left( \Phi^\dagger \eta \right)\left( \eta^\dagger \Phi \right) \nonumber \\[10pt]
 &+ \frac{\lambda_5}{2} \left[ \left(\Phi^\dagger \eta \right)^2 + \hc \right] + \frac{m_\Omega^2}{2} \, \text{Tr}(\Omega^\dagger \Omega)  
  + \frac{\lambda^{\Omega}_1}{2} \left( \Phi^\dagger \Phi \right) \text{Tr}\left( \Omega^\dagger \Omega\right) + \frac{\lambda^{\Omega}_2}{4} \, \text{Tr}(\Omega^\dagger \Omega )^2 \nonumber \\[10pt]
&+ \frac{\lambda^{\eta}}{2} \left( \eta^\dagger \eta \right) \text{Tr}\left( \Omega^\dagger \Omega\right) 
+ \mu_1 \, \Phi^\dagger \, \Omega \, \Phi + \mu_2 \, \eta^\dagger \, \Omega \, \eta \, . \label{eq:scpot}
\end{align}
%%%%%%%%%%%%%%%%%%%%%%%%%%%%%%%%%%%%%%%%%%%%%%%%%%%%
The spontaneous electroweak symmetry breaking will be driven by the neutral component of $\Phi$ and $\Omega$.
The field $\eta$ can not acquire a VEV due to the conservation of $\mathbb{Z}_2$ symmetry. The $SU(2)_L$ doublets $\Phi$, $\eta$ and $SU(2)_L$ triplet $\Sigma$, $\Omega$ can be written as follows  
%%%%%%%%%%%%%%%%%%%%%%%%%%%%%%%%%%%%%%%%%%%%%%%%%%%%%%
\begin{equation}
\Phi = \left( \begin{array}{c}
\phi^+ \\
\phi^0
\end{array} \right), \, \eta = \left( \begin{array}{c}
\eta^+ \\
\eta^0
\end{array} \right) \, , \,
\Sigma = \left( \begin{array}{cc}
\frac{\Sigma^0}{\sqrt{2}} & \Sigma^+ \\
\Sigma^- & -\frac{\Sigma^0}{\sqrt{2}}
\end{array} \right) \, \text{and} \quad \Omega = \left( \begin{array}{cc}
\frac{\Omega^0}{\sqrt{2}} & \Omega^+ \\
\Omega^- & -\frac{\Omega^0}{\sqrt{2}}
\end{array} \right) \, . \label{eq:triplets}
\end{equation}  
%%%%%%%%%%%%%%%%%%%%%%%%%%%%%%%%%%%%%%
As already mentioned, the spontaneous electroweak symmetry breaking will be driven by the neutral components of $\Phi$ and $\Omega$,
%%%%%%%%%%%%%%%%%%%%%%%%%%%%%%%%%%%%%%%%
\begin{equation}
\langle \phi^0 \rangle = \frac{v_\Phi}{\sqrt{2}} \, , \quad \langle \Omega^0 \rangle = v_\Omega \, , \quad \langle \eta^0 \rangle = 0 \, , \label{eq:vevs}
\end{equation}  
%%%%%%%%%%%%%%%%%%%%%%%%%%%%%%%%%%%%%%%%%%%%%

Minimization of the total potential $\mathcal{V}(\Phi,\Omega,\eta) $ leads to the relations: 
\begin{align}
 m_\Omega^2&=\displaystyle M_0^2-\frac{\lambda_1^\Omega v_\Phi^2}{2}-\frac{\lambda_2^\Omega v_\Omega^2}{2},\,\,\text{ with } M_0^2\equiv \frac{v_\Phi^2\mu_1}{2v_\Omega}.  \label{Tadpole1} \\
m_\Phi^2&=\displaystyle \frac{v_\Phi^2\lambda_1}{2}-\frac{\mu_1}{2}v_\Omega+\frac{\lambda_{1}^\Omega}{4}v_\Omega^2.
\label{Tadpole2}
\end{align}
In the limit $M_{0} \gg v_{\Phi} $, we can solve  \eqref{Tadpole1} for $v_{\Omega}$. Keeping terms of $\mathcal{O}(v_{\Phi}/M_{0})$ we get the small induced triplet vacuum expectation value : 
\begin{equation}
\boxed{
v_{\Omega} \approx \frac{\mu_1 v_{\Phi}^{2}}{2 M_{0}^{2}}}~. \label{V-triplet-approx}
\end{equation}  
providing a natural explanation for the smallness of the $v_{\Omega}$ consistent with the naive limits obtained in Sec.\ref{sec2-Wmass} for the CDF-II measurements.\\
%%%%%%%%%%%%%%%%%%%%%%%%%%%%%%%%%%%%%%%%%%%%%%%%%%%%%%%%%%%%%%%%%%%%%%%%%%%%%%%%%%%%%
After electroweak symmetry breaking there are three physical CP-even neutral fields and one CP-odd neutral field as one of them is absorbed by the $Z$ gauge boson. Also, there are three charged scalars, among them two are physical and one of them is absorbed by the $W$ gauge boson. The mass matrix of CP-even neutral scalars in the basis $(h^0, \Omega^0)$ reads as 
\begin{eqnarray}
\mathcal{M}_S^2 &=& \left(\begin{array}{cc}
\lambda_1 v_\Phi^2
& \lambda_1^\Omega v_\Omega v_\Phi - \mu_1 \frac{v_\Phi}{\sqrt{2}} \\
\lambda_1^\Omega v_\Omega v_\Phi - \mu_1 \frac{v_\Phi}{\sqrt{2}}
& 2 \lambda_2^\Omega v_\Omega^2 + \frac{\mu_1}{2\sqrt{2}} \frac{v_\Phi^2}{v_\Omega}
\end{array}\right) \equiv \left(\begin{array}{cc}
A  &  B \\
B  &  C \\
\end{array}\right)
\end{eqnarray}
%%%%%%%%%%%%%%%%%%%%%%%%%%%%%%%%%%%%%%%%%%%%%%%%%
with the mass eigenvalues given as
\begin{align}
m_{h,H}^2=\frac{1}{2}(A+C \mp \sqrt{(A-C)^2 + 4 B^2})
\end{align}
where we will identify the $h$ as SM-like Higgs boson discovered at LHC~\cite{ATLAS:2012yve,CMS:2012qbp}.
%%%%%%%%%%%%%%%%%%%%%%%%%%%%%%%%%%%%%%%%%%%%%%%%%%%%%%%%%%%%%%%%%%%%%%%%%%%%%%%%%%%%%%%%%%%%%%%%
The mass matrix for the charged scalars in the basis $(\phi^{\pm},\Omega^{\pm})$ is given as
\begin{eqnarray}
\mathcal{M}_{H^\pm}^2 &=& \left(\begin{array}{cc}
\sqrt{2}\mu_1 v_\Omega
& \mu_1 \frac{v_\Phi}{\sqrt{2}} \\
\mu_1 \frac{v_\Phi}{\sqrt{2}}
& \mu_1 \frac{v_\Phi^2}{2\sqrt{2}v_\Omega}
\end{array}\right) \, . \nonumber 
\end{eqnarray}
where the zero eigenvalue eigenstate corresponds to the would be Goldstone boson of the charged gauge boson $W^{\pm}$. The other is a physical charged scalar which has mass
\begin{align}
m_{H^\pm}^2=\frac{\mu_1 (v_\Phi^2+4 v_\Omega^2)}{2\sqrt{2}v_\Omega}
\end{align}
As the charged Goldstone boson is a linear combination of $\phi^{+}$ and $\Omega^{+}$, the VEV of $\Omega$ will contribute to the $W$ boson mass as follows, 
\begin{eqnarray}
M_W^2 &=& \frac{1}{4} \, g^2 \left( v_\Phi^2 + 4 \, v_\Omega^2 \right) \, , \label{eq:mW} 
\end{eqnarray}  
%%%%%%%%%%%%%%%%%%%%%%%%%%%%%%%%%%%%%%%%%%%%%%%%%%%%%%%%%%%%%%%%%%%%%%%%%%
 Finally the masses of $\eta$ doublet components can be written as follows
\begin{eqnarray}
m_{\eta_R}^2 &=& m_{\eta}^2 + \frac{1}{2}\left(\lambda_3 + \lambda_4 + \lambda_5 \right) v_\Phi^2 + \frac{1}{2}\lambda^\eta v_\Omega^2 - \frac{1}{\sqrt{2}} \, v_\Omega \, \mu_2 \, \\
m_{\eta_I}^2 &=& m_{\eta}^2 + \frac{1}{2}\left(\lambda_3 + \lambda_4 - \lambda_5 \right) v_\Phi^2 + \frac{1}{2}\lambda^\eta v_\Omega^2 - \frac{1}{\sqrt{2}} \, v_\Omega \, \mu_2 \, 
\end{eqnarray}
\begin{eqnarray}
m_{\eta^{\pm}}^2 &=& m_{\eta}^2 + \frac{1}{2}\lambda_3 v_\Phi^2 + \frac{1}{2}\lambda^\eta v_\Omega^2 + \frac{1}{\sqrt{2}} \, v_\Omega \, \mu_2 \, .
\end{eqnarray}
%%%%%%%%%%%%%%%%%%%%%%%%%%%%%%%%%%%%%%%%%%%%%%
For later convenience we define the ``mean mass'' $m^2_{\eta^0} = (m_{\eta_R}^2+m_{\eta_I}^2)/2$. The mass difference $m_{\eta_R}^2-m_{\eta_I}^2$ depends only on $\lambda_5$, which as we show later, is also responsible for the smallness of the neutrino mass scale. Thus, the smallness of the neutrino masses implies that $\lambda_5$ is quite small which in turn implies that $\eta_R$ and $\eta_I$ are almost mass degenerate. 
%%%%%%%%%%%%%%%%%%%%%%%%%%%%%%%%%%%%%%%%%%%%%%%%
\subsection*{\bf Neutrino masses}
%%%%%%%%%%%%%%%%%%%%%%%%%%%%%%%%%%%%%%%%%%%%%%%%

\begin{figure}[h!]
\includegraphics[scale=1]{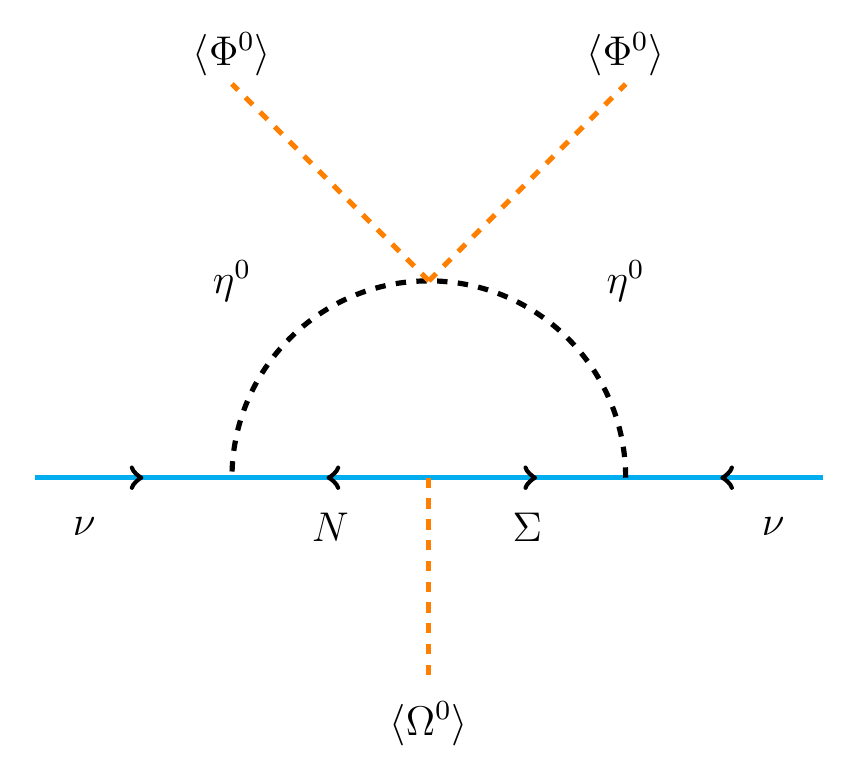}
\caption{One loop neutrino mass in the singlet-triplet scotogenic model, where $\eta^0=(\eta_R,\eta_I)$.}
\label{fig:scotoloop}
\end{figure}

The light neutrino masses are generated by the  singlet-triplet scotogenic loop as shown in Fig.~\ref{fig:scotoloop}, which can explain the small neutrino masses. The real triplet scalar whose VEV can account for the $W$ boson mass anomaly plays a crucial role in the case of neutrino mass generation.

Through the term proportional to the Yukawa coupling $Y_{\Omega}$ given in \eqref{eq:yukawa}, the triplet scalar generates a mixing between the singlet and triplet fermion fields $N$ and $\Sigma$. As a result, one gets the following tree level fermion masses
\begin{align}
m_{\chi^\pm}&=M_\Sigma\\
m_{\chi_1}&=\frac{1}{2}\left((M_\Sigma+M_N) - \sqrt{(M_\Sigma-M_N)^2+4 Y_\Omega^2 v_\Omega^2}\right)\\
m_{\chi_2}&=\frac{1}{2}\left((M_\Sigma+M_N)+\sqrt{(M_\Sigma-M_N)^2+4 Y_\Omega^2 v_\Omega^2}\right)
\end{align}
The mass eigenstates $\chi_{1,2}$ are determined by the $2\times 2$ orthogonal matrix $V(\alpha)$,
\begin{equation}
\left(\begin{array}{c}\chi_1\\ \chi_2\end{array}\right) = \left( \begin{array}{cc}
\cos \alpha & \sin \alpha \\
-\sin \alpha & \cos \alpha
\end{array} \right) \, \left(\begin{array}{c} \Sigma^0\\ N\end{array}\right) = V(\alpha)\left(\begin{array}{c} \Sigma^0\\ N\end{array}\right),
\end{equation}
with
\begin{equation}
\tan(2\alpha) = \frac{2 \, Y_\Omega v_\Omega}{M_\Sigma - M_N} \, .
\end{equation} 
%%%%%%%%%%%%%%%%%%%%%%%%%%%%%%%%%%%%%%%%%%%%%%%%%%%%%%%%%%%%%%%%%%%%%%
It is interesting to notice that due to the $\mathbb{Z}_2$ symmetry conservation, the lightest neutral eigenstate, $\chi_1$ or $\chi_2$ may also play the role of the dark matter~\cite{Hirsch:2013ola}.

%%%%%%%%%%%%%%%%%%%%%%%%%%%%%%%%%%%%%%%%%%%%%%%%%%%%%%%%%%%%%%%%%%

The expression for the neutrino mass matrix is
\begin{eqnarray}
(\mathcal{M}_\nu)_{\alpha\beta}&=&\sum_{\sigma=1}^2\left(\frac{ih_{\alpha \sigma}}{\sqrt{2}}\right)\left(\frac{-ih_{\beta \sigma}}{\sqrt{2}}\right) \frac{ m_{\chi_\sigma}}{(4\pi)^2} \left[\frac{m_{\eta_R}^2\ln\left(\frac{m_{\chi_\sigma}^2}{m_{\eta_R}^2}\right)}{m_{\chi_\sigma}^2-m_{\eta_R}^2}
-\frac{m_{\eta_I}^2\ln\left(\frac{m_{\chi_\sigma}^2}{m_{\eta_I}^2}\right)}{m_{\chi_\sigma}^2-m_{\eta_I}^2}\right] \label{eq:mnu}
\end{eqnarray}
where $h$ is a $3 \times 2$ matrix defined as 
\begin{equation}
h=\left(\begin{array}{c c} 
\frac{Y_\Sigma^1}{\sqrt{2}} & Y_N^1 \\
\frac{Y_\Sigma^2}{\sqrt{2}} & Y_N^2 \\
\frac{Y_\Sigma^3}{\sqrt{2}} & Y_N^3
\end{array}\right) \cdot V^T(\alpha) \, ,
\end{equation}
It proves convenient to write the neutrino mass matrix in
Eq.~\eqref{eq:mnu} as
\begin{equation} \label{eq:mnumat}
\mathcal{M}_\nu = h \, \Lambda \, h^T \, 
\end{equation}
where
\begin{equation}
\Lambda = \left( \begin{array}{cc}
\Lambda_1 & 0 \\
0 & \Lambda_2
\end{array} \right) \, , \quad \Lambda_\sigma = \frac{m_{\chi_\sigma}}{2 \, (4\pi)^2} \left[\frac{m_{\eta_R}^2\ln\left(\frac{m_{\chi_\sigma}^2}{m_{\eta_R}^2}\right)}{m_{\chi_\sigma}^2-m_{\eta_R}^2}
-\frac{m_{\eta_I}^2\ln\left(\frac{m_{\chi_\sigma}^2}{m_{\eta_I}^2}\right)}{m_{\chi_\sigma}^2-m_{\eta_I}^2}\right] \, .
\end{equation}
%%%%%%%%%%%%%%%%%%%%%%%%%%%%%%%%%%%%%%%
In order to compare with the current determination of neutrino oscillation parameters~\cite{10.5281/zenodo.4726908}, we will apply a Casas-Ibarra parametrization~\cite{Casas:2001sr} as follows
\begin{equation} \label{eq:CI}
h = U^\ast \, \sqrt{\widehat{\mathcal{M}}_\nu} \, R \, \sqrt{\Lambda}^{-1} \, .
\end{equation}
Here $R$ is a $3 \times 2$ complex matrix such that $R R^T =\mathbb{I}_{3}$, $\widehat{\mathcal{M}}_\nu$ is the diagonal neutrino mass matrix and $U$ is the leptonic mixing matrix.

%%%%%%%%%%%%%%%%%%%%%%%%%%%%%%%%%%%%%%%%%%%%%%%%%%%%%%%%%%%%%%%%%%%%%%%
\section{The $W$ boson mass and $S, T$ and $U$ parameters}
\label{sec4-STU}
%%%%%%%%%%%%%%%%%%%%%%%%%%%%%%%%%%%%%%%%%%%%%%%%%%%%%%%%%%%%%%

In the SM, the $W$ boson mass can be calculated very precisely in terms of the precisely measured input parameters $\{G_F, \alpha_{\text{em}}, M_Z, m_h, m_t, \alpha_s(M_Z)\}$. The $W$ boson mass is related with these parameters in the following way:
%%%%%%%%%%%%%%%%%%%%%%%%%%%%%%%%%
\begin{align}
M_W=\frac{M_Z}{2}\left(1+\sqrt{1-\frac{4\pi\alpha_{\rm em}}{\sqrt{2}G_F M_Z^2(1-\Delta r)}}\right),
\label{eq:MW}
\end{align}
where $\Delta r$ represents the quantum corrections. Note that here the $M_W$ and $M_Z$ are the renormalized masses in the on-shell scheme. Taking the central values of the input parameters, $M_Z = 91.1876$ GeV, $\alpha_{\rm em}^{-1}(0) = 137.036$, $s_W^2 = 0.2315$, $M_t = 172.76$ GeV, $M_h = 125. 25$ GeV, and $G_F = 1.1663787 \times 10^{-5}$ GeV$^{-2}$, one finds from Eq. (\ref{eq:MW}) $M_W= 80.354$ GeV, which is 7-$\sigma$ below the CDF-II measurement~\cite{CDF:2022hxs}.\\
%%%%%%%%%%%%%%%%%%%%%%%%%%%%%%%%%%%%%%%%%%%%%%%%%%%%%%%%%%%%%%%%%%%%%%%%%%%%%%%%%%%%%%%%%%%%%%%%
 Provided that the new physics mass scale is higher than the electroweak scale, and that it contributes only through virtual loops to the electroweak precision observables, the dominant BSM effects can be parametrised by three gauge boson self-energy parameters named the oblique parameters $S, T$ and $U$. These can be considered as the reparametrisations of the variables $\Delta\rho$, $\Delta\kappa$ and $\Delta r$, which absorb the radiative corrections to the total $Z$ coupling strength, the effective weak mixing angle, and the $W$ mass, respectively. For a given new physics model, the $S$, $T$, $U$ predictions consist of the sum of the BSM contributions and the non-vanishing SM remainders when the $m_h$ and $m_t$ values differ from those used for the SM reference. The deviation of $M_W^{\text{CDF}}$ from its SM prediction can be parameterized in terms of the oblique parameters, $S, T$ and $U$ as follows: 
%%%%%%%%%%%%%%%%%%%%%%%%%%%%%%%%%%%%%%%%%%%%%%%%
\begin{align}
M_W =M_W^{\rm SM}\Big[\frac{\sqrt{v_\Phi^2 + 4 v_\Omega^2}}{v_\Phi}-\frac{\alpha_{\rm em}}{4(c_W^2-s_W^2)}(S-1.55T-1.24 U)\Big] 
\end{align}
where the first term comes from the tree level contribution of the triplet scalar VEV while all BSM particles contribute through loops to the $S$, $T$, $U$ parameters\footnote{In a given model, the contribution of different BSM particles to $S$, $T$, $U$ parameters need not be of equal importance. }.

%%%%%%%%%%%%%%%%%%%%%%%
Recently, Ref.~\cite{Lu:2022bgw} gave the values of these parameters from an analysis of precision electroweak data including the CDF-II new result of the $W$-mass:
%%%%%%%%%%%%%%%%%%%%%%%
\begin{align}
S=0.06\pm 0.10,\,\,  T=0.11\pm 0.12 \,\,\,\,\text{and}\,\,\,\, U=0.14\pm 0.09,
\end{align}
with the correlation 
\begin{align}
\rho_{ST}=0.90,\,\,  \rho_{SU}=-0.59 \,\,\,\,\text{and}\,\,\,\, \rho_{TU}=-0.85.
\end{align}
%%%%%%%%%%%%%%%%%%%%%%%%%%%%
On the other hand, since the values of the $U$ parameter are found to be very small in many new physics models, it is reasonable to consider $U\approx 0$. With this assumption, Ref.~\cite{Lu:2022bgw} found the following values of $S$ and $T$:
%%%%%%%%%%%%%%%%%%%%%%%%%%%%%%%%%%%%%
\begin{align}
S=0.15\pm 0.08 \,\,\,\text{and}\,\,\, T=0.27\pm 0.06\,\,\text{with the correlation}\,\, \rho_{ST}=0.93
\end{align}

%%%%%%%%%%%%%%%%%%%%%%%%%%%%%%%%%%%%%%%%%%%%%%%%%%%%%%%%%%%%%%%%%%%%%%%%%%%
\subsection{$S$, $T$, $U$ parameters in singlet-triplet scotogenic model}
%%%%%%%%%%%%%%%%%%%%%%%%%%%%%%%%%%%%%%%%%%%%%%%%%%%%%%%%%%%%%%%%%%%%%%%%%%
Since the scalar triplet~($\Omega$), scalar doublet~($\eta$) and triplet fermion~($\Sigma$) couple to the $\mathrm{SU(2)_L}$ gauge bosons, they can potentially  affect the the oblique parameters $S$, $T$, $U$ through loop corrections. 
The contributions to $S$, $T$, $U$ are given by \cite{Dedes_2014,Khan:2016,Abada:2018}
%%%%%%%%%%%%%%%%%%%%%%%%%%%%%%%%%%%%%%%%%%%%%%%%%%%%%%%%%%%%%%%%%%%%%%%%%
\begin{align}	\label{eq:theo}
	S & \simeq  \underbrace{\frac{1}{12 \pi} \log \left(\frac{m^2_{\eta^0}}{m^2_{\eta^{+}}}\right)}_{\text{Scalar doublet contribution}} + \frac{1}{18\pi}
	\nonumber\\
	T & \simeq 	\underbrace{\frac{1}{6\pi}\frac{1}{\sin^2(\theta_W) \cos^2(\theta_W)} \frac{\Delta M}{M_Z^2}}_{\text{Scalar triplet contribution}} + 
 \underbrace{\frac{2\sqrt{2}G_F}{(4\pi)^2\alpha_{em}} \left[ \frac{m^2_{\eta^0} + m^2_{\eta^{+}}}{2} - \frac{•m^2_{\eta^0}m^2_{\eta^+}}{m^2_{\eta^+} - m^2_{\eta^0}}\log \left( \frac{m^2_{\eta^+}}{m^2_{\eta^0}}\right) \right]}_{\text{Scalar doublet contribution}}
	\nonumber \\
	U & \simeq \underbrace{\frac{\Delta M}{3\pi M_H^{\pm}}}_{\text{Scalar triplet contribution}}
	\\\nonumber
	& + \underbrace{\frac{1}{12\pi} \left[ - \frac{5 m_{\eta^+}^4 - 22 m_{\eta^+}^2 m_{\eta^0}^2 + 5 m_{\eta^0}^4}{3 \left( m_{\eta^+}^2 - m_{\eta^0}^2 \right)^2 } + \frac{\left( m_{\eta^+}^2 + m_{\eta^0}^2 \right) \left(m_{\eta^+}^4 - 4 m_{\eta^+}^2 m_{\eta^0}^2 + m_{\eta^0}^4 \right)}{\left( m_{\eta^+}^2 - m_{\eta^0}^2 \right)^3}
	\right]}_{\text{Scalar doublet contribution}}
\end{align}
%%%%%%%%%%%%%%%%%%%%%%%%%%%%%%%%%%%%%%
where $\Delta M = M_{\Omega^{\pm}} - M_{\Omega^0}$ corresponds to the mass splitting between the charged and neutral components of the triplet scalar $\Omega$. This mass splitting is directly proportional to the triplet VEV $v_\Omega$. Due to the smallness of the triplet VEV, $\Delta M << M_Z$. This in turn implies that the triplet scalar contribution to the $S$, $T$, $U$ parameters will be negligibly small compared to the contribution coming from the doublet scalar $\eta$, a fact which we have also verified numerically. Furthermore, since the triplet fermion $\Sigma$ has an invariant Majorana mass term, its mass is independent of the electroweak scale and is typically taken to be much larger than the electroweak scale. Thus, in writing the above relations, we have ignored the contribution of the $\Sigma$ as its contribution is expected to be negligibly small. Finally, the SM gauge singlet fermion $N$ also has a large invariant mass and in addition, being a gauge singlet it obviously does not contribute to the $S$, $T$, $U$ parameters.

Thus, in the end there are primarily two different important contributions to the mass of the $W$ boson in our model:
\begin{enumerate}
 \item The direct tree level contribution of the triplet scalar though its VEV $v_\Omega$.
 \item The loop level contribution of the doublet scalar $\eta$.
\end{enumerate}

\begin{figure}[!htbp]
	\centering
	\includegraphics[scale=0.7]{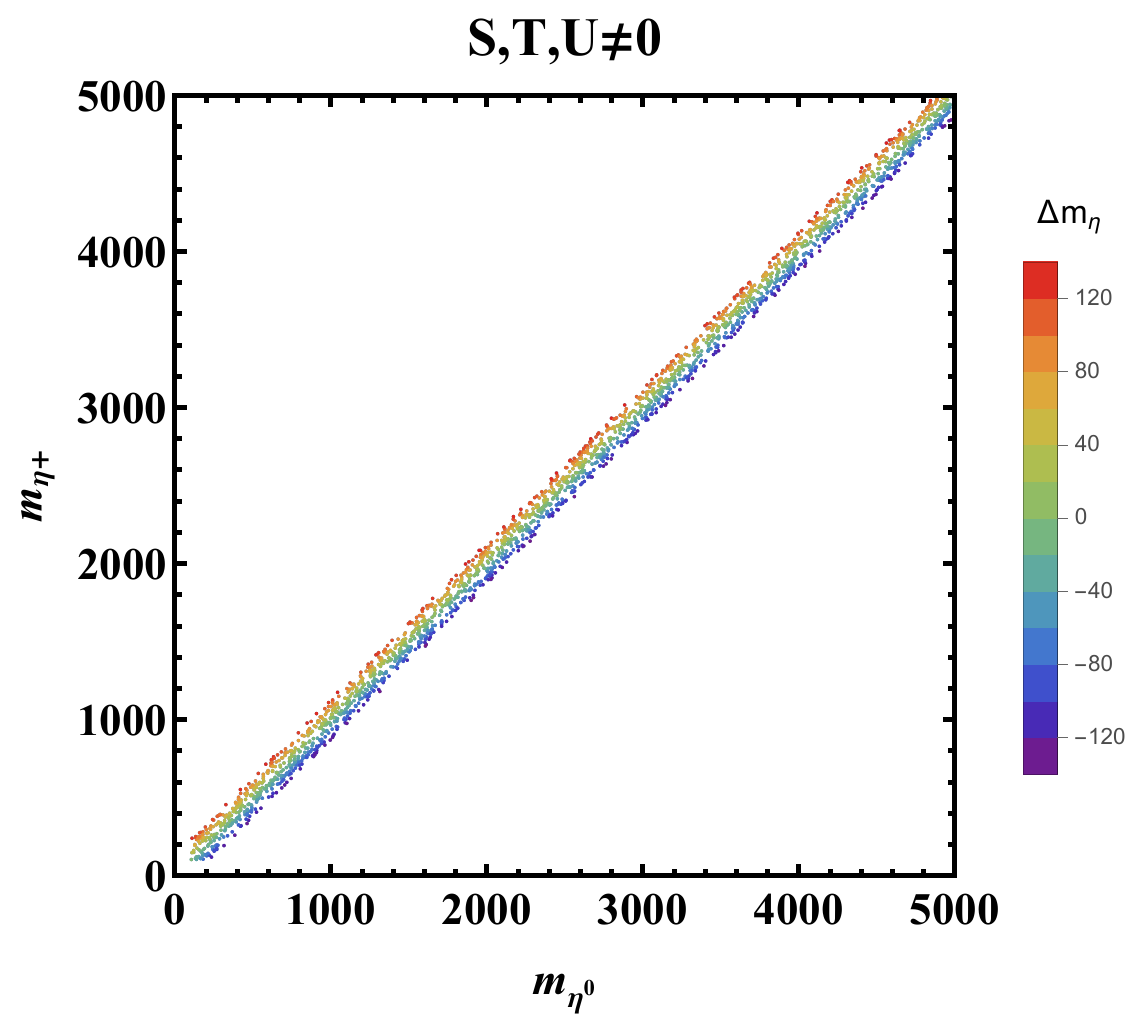}
		\includegraphics[scale=0.7]{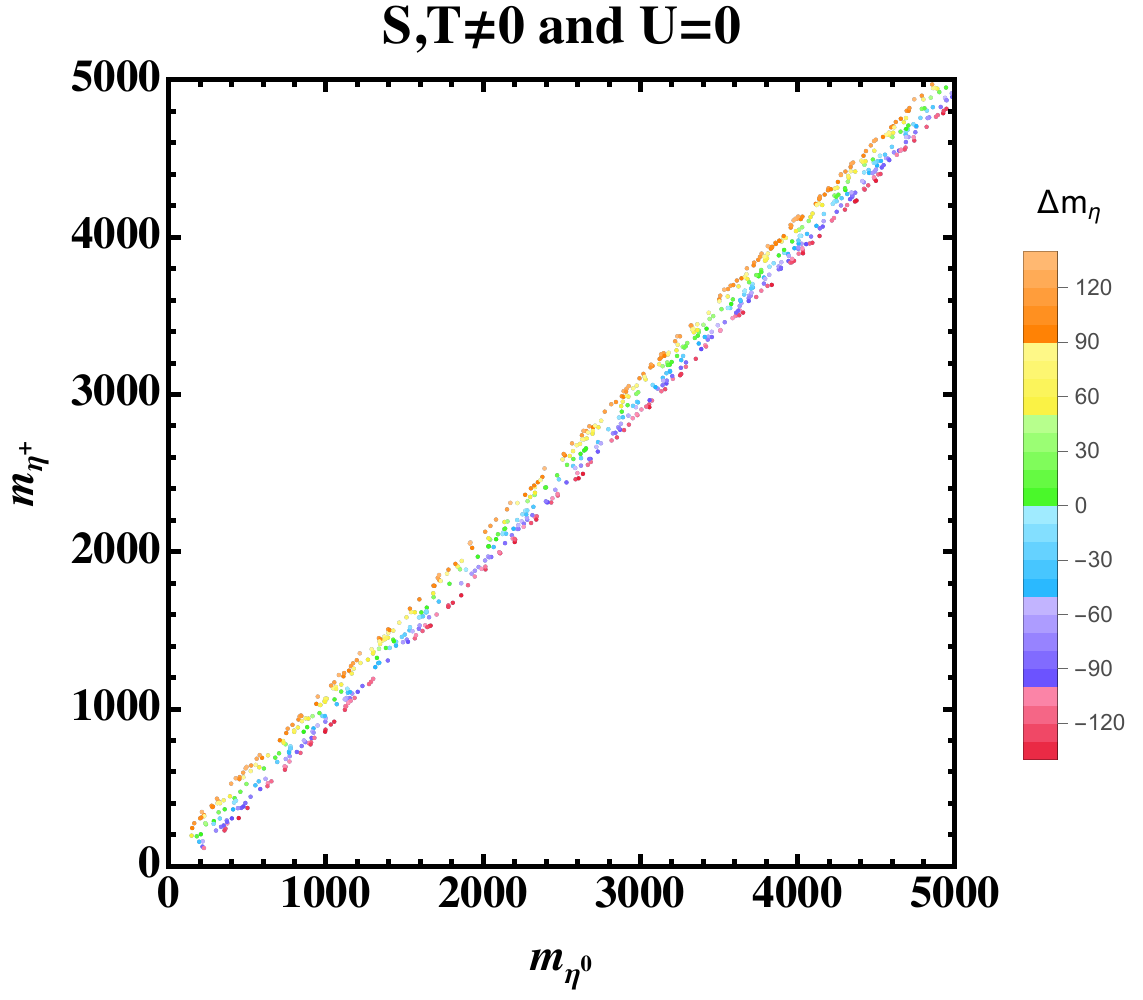}
	\caption{Values of $m_{\eta^0}$ vs $m_{\eta^+}$ with the mass splitting $\Delta m_\eta$ allowed by the new $S$, $T$, $U$ parameter space as given in \cite{Lu:2022bgw}. The left panel is with $S$, $T$, $U\neq 0$ and the right panel is for $S$, $T\neq 0$, $U=0$. See text for details.}
	\label{fig:etavsetap}
\end{figure}

Being two independent sources, in our model both or only one of them can be the dominant contribution leading to the $W$ boson mass in agreement with CDF-II measurement. In previous section we have already discussed the scenario where the  triplet scalar contribution is the dominant contribution. Here, we want to explore the scenario where $v_\Omega$ is much smaller than the GeV scale so that its contribution to $W$ mass is negligible. In such a case, the BSM contribution to $W$ mass is purely coming from quantum loop corrections, i.e, only through the $S$, $T$ and $U$ parameters.
In such a case the CDF-II measurements imply a strong correlation between the masses of the charged and neutral components of the doublet scalar $\eta$ as shown in Fig.\ref{fig:etavsetap} with the color bar indicating the mass splitting $\Delta m_{\eta} = m_{\eta^+}$ - $m_{\eta^0}$ between the charged and neutral components .
%%%%%%%%%%%%%%%%%%%%%%%%%%%%%%%%%%%%%%%%%%%%%%%%%%%%%%%%%%%%%%%%%%%%%%%%%%%%%%%%%%%%

 As mentioned above, the dominant contributions to $W$ mass in this case is assumed to come from the doublet scalar. This allows us to constrain the parameter space of allowed values of masses of $m_{\eta^0}$ and $m_{\eta^+}$ that can simultaneously satisfy the $S$, $T$ and $U$ constraints for the new CDF-II data~\cite{Lu:2022bgw}. Since the contribution of doublet scalar to $S$, $T$, $U$ are highly sensitive to the mass and mass differences of the charged and neutral components of the doublets as can be seen in \eqref{eq:theo}, the allowed parameter space is a narrow band as shown in \ref{fig:etavsetap}. The narrow bands are also color coded with respect to the mass differences of the neutral and charged components of doublet scalar to better understand the sensitivity to mass differences. We can infer a maximum allowed value for the mass difference $\Delta m_{\eta}$ to be around 120 GeV to satisfy the $S$, $T$ and $U$ constraints. 
%%%%%%%%%%%%%%%%%%%%%%%%%%%%%%%%%%%%%%%
\section{Dark matter constraints}
\label{sec5-const}
%%%%%%%%%%%%%%%%%%%%%%%%%%%%%%%%%%%%%%%%%
In this section we collect the results of our dark matter analysis. Since both $N$ and $\Sigma$ fermions have invariant mass terms which can be much larger than the electroweak scale, we have assumed them to be heavy. With this assumption, the  $\mathbb{Z}_2$ symmetry conservation makes the lightest of the two scalar eigenstates $\eta_R$ and $\eta_I$ a viable dark matter candidate. In our analysis we take $\eta_R$ as the dark matter candidate with $\lambda_5<0$~(the opposite scenario with $\lambda_5>0$ would have $\eta_I$ as the dark matter candidate). We implemented the model in SARAH~\cite{Staub:2015kfa} and SPheno~\cite{Porod:2011nf} to calculate all the vertices, mass matrices, tadpole equations, whereas the thermal component of the dark matter relic, as well as dark matter direct detection cross sections are determined by micrOMEGAS-5.0.8~\cite{Belanger:2014vza}.\\
%%%%%%%%%%%%%%%%%%%%%%%%%%%%%%%%%%%%%%%%%%%%%%%%%%%%%%%%
\begin{table}[!htbp]
\begin{center}
\begin{tabular}{ | >{\centering\arraybackslash}m{1in} |>{\centering\arraybackslash}m{3in}| }
 \hline
 \multicolumn{2}{|c|}{Scalar Dark matter - $\eta_R$} \\
 \hline \hline
 Parameter & Range \\
\hline
&\\[-12pt]
$m_{\eta}^2$  & $\bm{[} \, 10^2, 2.5\times10^7 \, \bm{]} \; \text{GeV}$ \\
$M_{\Sigma}, M_N$  & $\bm{[} \, m_{\eta_R} + 10, m_{\eta_R} + 5000 \, \bm{]} \; \text{GeV}$ \\ 
$\mu_i$  & $\bm{[} \,10, 3000 \, \bm{]} \; \text{GeV}$ \\
$v_{\Omega}$  & $\bm{[} \, 2, 6 \, \bm{]} \; \text{GeV}$ \\
$\lambda_5$  & $\bm{[} \, -10^{-5}, -10^{-1} \, \bm{]} $ \\ 
$\lambda_4$  & $\bm{[} \, -\lambda_5
-10^{-5}, -\lambda_5-10^{-1} \, \bm{]} $ \\ 
$\lambda_2,\lambda_3$  & $\bm{[} \, 10^{-5}, 10^{-1} \, \bm{]} $ \\ 
$\lambda_i^\Omega$  & $\bm{[} \, 10^{-5}, 10^{-1} \, \bm{]} $ \\
$\lambda^\eta$  & $\bm{[} \, 10^{-5}, 10^{-1} \, \bm{]} $ \\ 
$Y_{\Omega} $  & $\bm{[} \, 10^{-5}, 10^{-1} \, \bm{]} $ \\ 
$Y_{\nu} $  & $\bm{[} \, 10^{-5}, 10^{-1} \, \bm{]} $ \\ 
\hline
\end{tabular}
\end{center}
\caption{Value range for the numerical parameter scan for a scalar dark matter candidate.}
  \label{tab:scalar-dark-matter}
\end{table}
%%%%%%%%%%%%%%%%%%%%%%%%%%%%%%%%%%%%%%%%%%%%%%%%%%%%%

\begin{figure}[!htbp]
\centering
\includegraphics[scale=1.5]{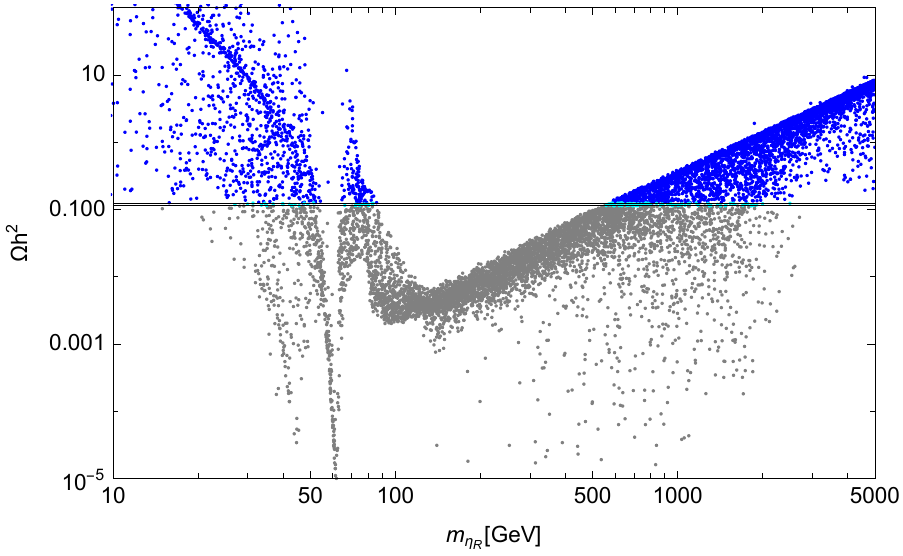}
\caption{Relic density versus the mass of the dark matter candidate $m_{\eta_R}$ for various input parameters as given in Table.~\ref{tab:scalar-dark-matter}. The cyan points are within the 3-$\sigma$ range of observed relic density~\cite{Planck:2018vyg}, whereas the blue and gray points correspond to over-abundant and under-abundant dark matter respectively.}
\label{fig:SDM-relic}
\end{figure}

We show in Fig.~\ref{fig:SDM-relic}, the behaviour of relic density as a function of the mass of the scalar dark matter candidate $\eta_R$. The narrow band is the 3-$\sigma$ allowed range for relic density~\cite{Planck:2018vyg}: $0.1126 \leq \Omega_{\eta_R}h^2\leq 0.1246$. Our numerical scan was performed varying the input parameters as given in Table.~\ref{tab:scalar-dark-matter}, assuming logarithmic steps. The range of $ m_{\eta} $ is taken to be 10 to 5000 GeV. The mass differences $ m_\Sigma$ - $m_{\eta_R} $ and $ m_N$ - $m_{\eta_R} $ each have a range of 10 to 5000 GeV. $ \lambda_1 $ is fixed in such a way that the lighter neutral CP-even scalar $ h $ has a mass of $ 125.25 \pm 0.17 $ GeV~\cite{Zyla:2020zbs}. The condition $ \lambda_4 + \lambda_{5} < 0 $ is imposed such that $ m_{\eta_R} < m_{\eta^\pm} $. The meaning of color code in Fig.~\ref{fig:SDM-relic} is as follows: the cyan points are within the 3-$\sigma$ range, whereas the blue and gray points are above and below the 3-$\sigma$ range respectively.

The reasons for the various dips can be understood by looking in detail into the $\eta_R$ annihilation and co-annihilation channels given in Fig.~\ref{fig:Annihilation} in Appendix~\ref{app2-feynman}. The first dip at $m_{\eta_R}\sim M_Z/2$ is due to annihilation and co-annihilation via s-channel $Z$-exchange. The second dip around $m_{\eta_R} \sim 60$ GeV corresponds to efficient annihilations via s-channel Higgs exchange. Note that some of the points in this dip can also be ruled out from the current collider limits on $\text{BR}(h \to \text{inv})$. The Higgs-exchange dip is more efficient than the $Z$-mediated one as $Z$-mediated channel is momentum suppressed. For heavier $\eta_R$ masses, quartic interactions with gauge bosons starts contributing. For example, with $m_{\eta_R} \geq 80$ GeV, annihilations of $\eta_R$ into $W^+W^-$ and $ZZ$ via quartic couplings are particularly important, hence the third drop in the relic abundance. For $m_{\eta_R} \geq 120$ GeV and $m_{\eta_R} \geq m_t$ GeV, $\eta_R$ can annihilate also into two Higgs bosons, $hh$ and $t\bar{t}$, respectively.  For large $m_{\eta_R}$, the relic density increases due to the suppressed annihilation cross section, which drops as $\sim 1/m_{\eta_R}^2$. Notice also that when $\lambda_5$ is small, coannihilation channels with both $\eta_I$ and $\eta^{\pm}$ may occur in all regions of the parameter space.\\

%%%%%%%%%%%%%%%%%%%%%%%%%%%%%%%%%%%%%%%%%%%%%%%%%%%%%%%%%%%%%
\begin{figure}[!htbp]
\centering
\includegraphics[scale=1.5]{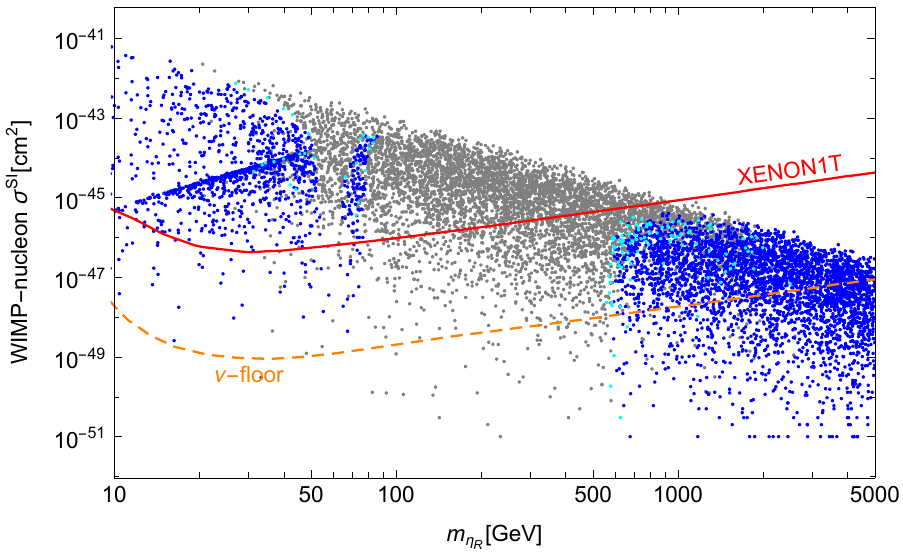}
\caption{Spin-independent WIMP-nucleon cross section versus the mass of the dark matter candidate $m_{\eta_R}$ for various input parameters as given in Table.~\ref{tab:scalar-dark-matter}. The color scheme is the same as in Fig.~\ref{fig:SDM-relic}. The red line denotes the latest upper bound from the XENON1T collaboration~\cite{XENON:2018voc} and the dashed orange line corresponds to the “neutrino floor” lower limit~\cite{Billard:2013qya,Billard:2021uyg}.}
\label{fig:DDscalar}
\end{figure}
%%%%%%%%%%%%%%%%%%%%%%%%%%

The tree-level spin-independent $\eta_R$-nucleon cross section is mediated by the Higgs and the $Z$-portals as seen in Fig.~\ref{fig:direct} in Appendix~\ref{app2-feynman}. For non-zero $\lambda_5$, there is a small mass splitting between $\eta_I$ and $\eta_R$, so the interaction through the $Z$-boson is kinematically forbidden or leads to inelastic scattering, hence can be neglected compared to the Higgs-exchange contribution.

In Fig.~\ref{fig:DDscalar}, we show the the direct detection prospects of our dark matter $\eta_R$,  for the range of parameters covered by our scan given in Table.~\ref{tab:scalar-dark-matter}. The color code of displayed points is the same as in Fig.~\ref{fig:SDM-relic}. The red line denotes the latest upper bound from the XENON1T collaboration~\cite{XENON:2018voc}. There are constraints from other experiments as well, such as LUX~\cite{LUX-ZEPLIN:2018poe} and PandaX-II~\cite{PandaX:2018wtu}, but weaker when compared to the XENON1T limit. We also show the lower limit corresponding to the “neutrino floor” from coherent elastic neutrino scattering~\cite{Billard:2013qya,Billard:2021uyg}. We see from Fig.~\ref{fig:DDscalar} that all the low-mass solutions with the correct dark matter relic density are ruled out by the XENON1T direct detection cross section upper limits.

Although we have taken the dark sector fermions of the model to be heavier than the dark scalars but this need not be the case. The fermionic dark matter in this model is also very interesting. In this case the relic abundance of $\Sigma^0$ is determined by the annihiliation and co-annihiliation of itself and $\Sigma^{\pm}$. These processes force  the triplet dark matter mass to be in the range 2.3 TeV to 2.4 TeV. Also the direct detection only occurs through a loop. On the other hand, for pure singlet $N$, the dark matter mass can be much smaller compared to the pure triplet case. The main signature of $N$ is the annihilation into neutrinos and charged leptons.  Again, the disadvantage is that direct detection occurs only at the one loop level.  Finally, note that the nonzero VEV of the scalar triplet $\Omega$ induces a mixing between $\Sigma^0$ and $N$. If the Yuakwa coupling between $N$ and $\Sigma$ through $\Omega$ is large, then the  fermionic dark matter candidate $\chi_i$ can a mixed state between $N$ and $\Sigma^0$. In comparison to models with only singlets or triplets, this interaction results in an enhanced dark matter phenomenology.  This same mixing can also give tree-level direct detection. Unlike the pure singlet or pure triplet case, the singlet-triplet mixed dark matter has the best features of singlet-only or triplet-only scenarios with a better prospect of direct detection. For a detailed discussions on fermionic dark matter, we refer to~\cite{Hirsch:2013ola}.

%%%%%%%%%%%%%%%%%%%%%%%%%%%%%%%%%%%%%%%%%%%%
\section{Conclusions}
\label{sec6-conclusions}
%%%%%%%%%%%%%%%%%%%%%%%%%%%%%%%%%%%%%%%%%%%
We have presented a singlet-triplet scotogenic dark matter model to generate tiny neutrino masses at one loop level. In this model, the dark sector particles run in the loop generating the neutrino masses. The particle content of the model is simple. Apart from SM particles the model contains the SM gauge singlet fermion $N$ and well as the hypercharge zero $SU(2)_L$ triplet fermion $\Sigma$ along with the  $SU(2)_L$ doublet scalar $\eta$. Additionally a hypercharge zero $SU(2)_L$ triplet scalar $\Omega$ is needed in order to complete the neutrino mass loop diagram. We looked at the prospects of the model satisfying the recent $W$ boson mass measurements by the CDF-II collaboration. In the model there are two main corrections to the SM value of $W$ mass, namely, direct modification due to the VEV $v_\Omega$ of the triplet scalar and loop corrections due to presence of the dark scalar $\eta$. We have analyzed both these possibilities. We find that the CDF-II result can be explained in both scenarios. We derived the resulting constraints on the VEV of the triplet scalar as well as the mass spectra of the doublet scalar components. Finally we looked at the dark matter phenomenology in the model and showed that the dark matter in our model indeed satisfies both  the relic abundance and direct detection constraints over a large range of the parameter space.  
%%%%%%%%%%%%%%%%%%%%%%%%%%%%%%%%%%%%%%%%%%%%%%%%%%%%%%
\begin{acknowledgments}
The work of RS is supported by the Government of India, SERB Startup Grant SRG/2020/002303. The work of S.M. is supported by KIAS Individual Grants (PG086001) at Korea Institute for Advanced Study.
\end{acknowledgments}

\appendix

\section{Definition of $S$, $T$ and $U$ parameters}
%%%%%%%%%%%%%%%%%%%%%%%%%%%%%%%%%%%%%%%%%%%%%%%%%%%
Following closely the notation by Peskin and Takeuchi in \cite{Peskin:1991sw}, the $S, T$ and $U$ parameters can be defined as
%%%%%%%%%%%%%%
\begin{subequations}
\begin{align}
\alpha \, S &\equiv 4 e^{2} \: \frac{d}{dp^{2}} \left [\Pi_{33}(p^{2}) - \Pi_{3Q}(p^{2})
\right ] \biggl |_{p^{2}=0} \;, \\
\alpha \, T &\equiv \frac{e^{2}}{s_{W}^{2} c_{W}^{2} m_{Z}^{2}}\: 
\left [\Pi_{11}(0) - \Pi_{33}(0) \right ]\;, \\
\alpha \, U &\equiv 4 e^{2} \: \frac{d}{dp^{2}} \left [\Pi_{11}(p^{2}) - \Pi_{33}(p^{2})
\right ]\biggl |_{p^{2}=0} \;,  
\end{align}
\label{eq:STU}
\end{subequations}
%%%%%%%%%%%%%%%%
where $\alpha = e^{2}/4\pi$. $\Pi_{IJ}\equiv \Pi_{IJ}(p^{2})$ are the various vaccum polarisation diagrams
where $I$ and $J$ may be photon ($\gamma)$, $W$ or $Z$, 
%%%%%%%%%%%%%%%%%%
\begin{subequations}
\begin{align}
\Pi_{\gamma\gamma} &= e^{2}\: \Pi_{QQ} \;, \\
\Pi_{Z\gamma} &= \frac{e^{2}}{c_{W}s_{W}} \: \left ( \Pi_{3Q} - s^{2} \Pi_{QQ} \right )\;,\\
\Pi_{ZZ} &= \frac{e^{2}}{c_{W}^{2} s_{W}^{2}} \: 
\left (\Pi_{33} - 2 s^{2} \Pi_{3Q} + s^{4} \Pi_{QQ}\right )\;,\\
\Pi_{WW} &= \frac{e^{2}}{s_{W}^{2}} \: \Pi_{11} \;,
\end{align}
\end{subequations}
%%%%%%%%%%%%%%%%% 
where $s_{W}=\sin\theta_{W}, c_{W}=\cos\theta_{W}$.

\section{Feynman diagrams for dark matter relic density and direct detection}
\label{app2-feynman}
\begin{figure}[!htbp]
	\centering
	\includegraphics{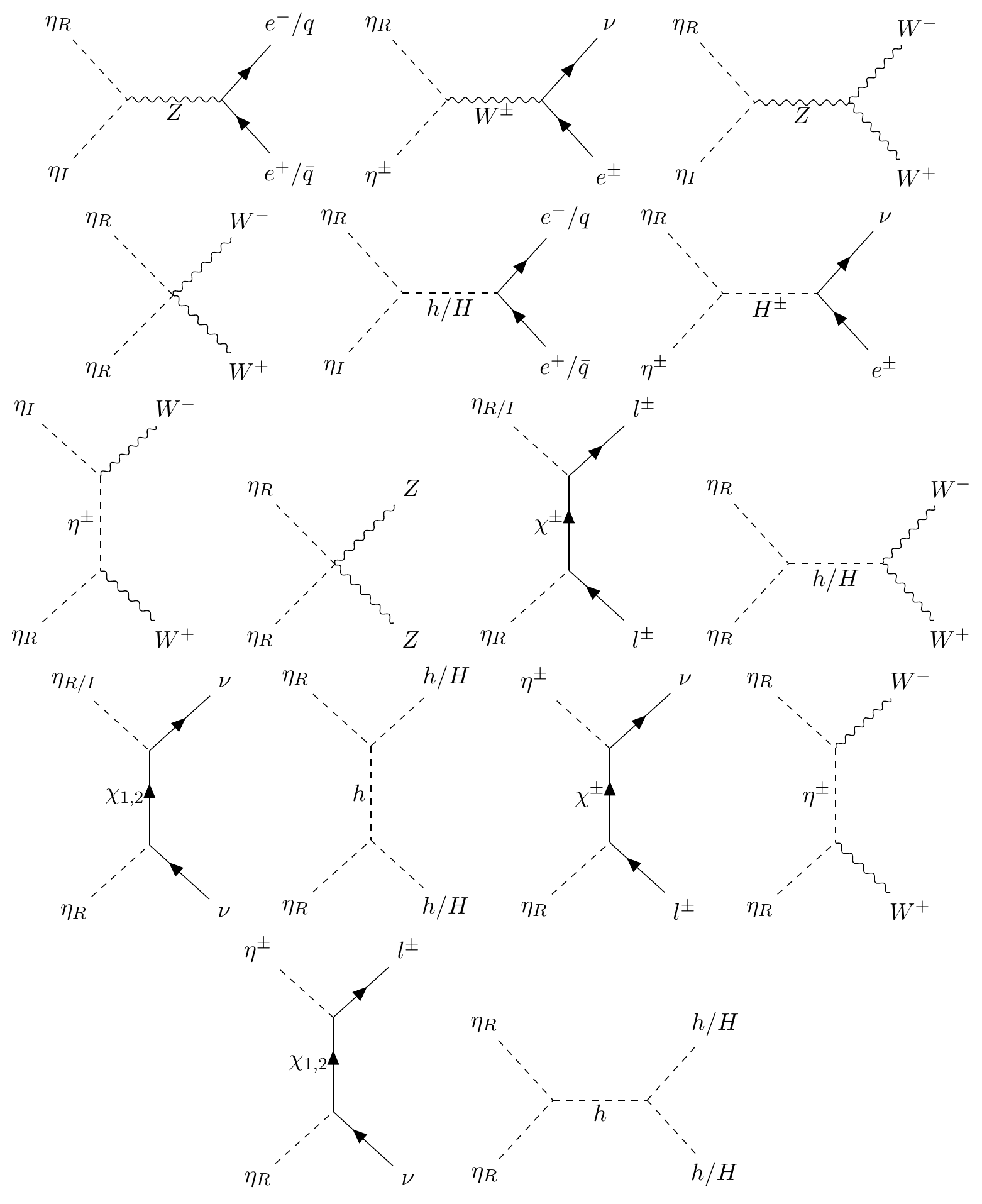}
	\caption{Annihilation and co-annihilation Feynman diagrams that contribute to the relic density of $ \eta_R $.}
	\label{fig:Annihilation}
\end{figure}

\begin{figure}[!htbp]
	\centering
	\includegraphics{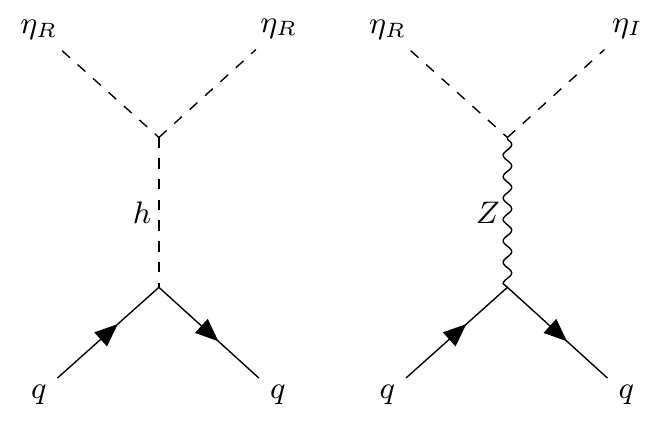}
	\caption{$ \eta_R $-nucleon scattering Feynman diagrams relevant for direct detection.}
	\label{fig:direct}
\end{figure}

%%%%%%%%%%%%%%%%%%%%%%
We show some of the most important Feynman diagrams for determining the cosmological relic density, assuming $\eta^R$ is the dark matter. The primary dark matter annihilation and coannihilation channels are depicted in Fig.~\ref{fig:Annihilation}. Coannihilations with both $\eta^I$ and $\eta^{\pm}$ are feasible in addition to the standard $s$-wave annihilation into quarks and gauge bosons, which is mediated by the SM like Higgs boson. The $Z$ boson, as well as the new fermions, can mediate these interactions. At the tree level, the diagrams in Fig.~\ref{fig:direct} contribute to the spin-independent $\eta^R$-nucleon elastic scattering cross section. Only when the separation between the masses of $\eta^R$ and $\eta^I$ is minimal~(small $\lambda_5$ values) diagram on the right matter and leads to inelastic cross section.
%
%%%%%%%%%%%%%%%%%%%%%%%%%%%%%%%%%%%%%%%%%%%%%%%%%%%%%%%%%%%%%%
%%%%%%%%%%%%%%%%%%%%%%%%%%%%%%%%%%%%%%
%
%%%%%%%%%%%%%%%%%%%%%%%%%%%%%%%%%%%%%%
%%%%%%%%%%%%%%%%%%%%%%%%%%%%%%%%%%%%%%%%%%%%%%%%%%%%
\newpage

\bibliographystyle{utphys}
\bibliography{bibliography} 
\end{document}